\documentclass[aps,prl,superscriptaddress,twocolumn,showpacs,intlimits,amsmath,amssymb,floats,longbibliography]{revtex4-1}

\usepackage{times}
\usepackage{epsfig}
\usepackage{graphicx}
\usepackage{pstricks}
\usepackage{dcolumn}		
\usepackage{bm}			
\usepackage{ulem}
\usepackage{units}		
\usepackage{color}
\usepackage[english]{babel}
\usepackage[permil]{overpic}
\usepackage{float}
\usepackage[colorlinks=true,allcolors=blue]{hyperref}  

\def\ligand{\underline{L}}
\def\R{{\it R}}

\begin{document}



\title{Electronic Structure Trends Across the Rare-Earth Series \\in Superconducting Infinite Layer Nickelates}

\author{Emily Been}
\affiliation{Stanford Institute for Materials and Energy Sciences, SLAC National Accelerator Laboratory, 2575 Sand Hill Road, Menlo Park, CA 94025.}
\affiliation{Department of Physics, Stanford University, Stanford, CA 94305.}
\author{Wei-Sheng Lee}
\affiliation{Stanford Institute for Materials and Energy Sciences, SLAC National Accelerator Laboratory, 2575 Sand Hill Road, Menlo Park, CA 94025.}
\author{Harold Y. Hwang}
\affiliation{Stanford Institute for Materials and Energy Sciences, SLAC National Accelerator Laboratory, 2575 Sand Hill Road, Menlo Park, CA 94025.}
\affiliation{Department of Applied Physics, Stanford University, Stanford, CA 94305.}
\author{Yi Cui}
\affiliation{Department of Materials Science and Engineering, Stanford University, Stanford, CA 94305.}
\author{Jan Zaanen}
\affiliation{Institute Lorentz for Theoretical Physics, Leiden University, 2300 RA Leiden, Netherlands}
\author{Thomas Devereaux}
\affiliation{Stanford Institute for Materials and Energy Sciences, SLAC National Accelerator Laboratory, 2575 Sand Hill Road, Menlo Park, CA 94025.}
\affiliation{Department of Materials Science and Engineering, Stanford University, Stanford, CA 94305.}
\author{Brian Moritz}
\affiliation{Stanford Institute for Materials and Energy Sciences, SLAC National Accelerator Laboratory, 2575 Sand Hill Road, Menlo Park, CA 94025.}
\author{Chunjing Jia} 
\affiliation{Stanford Institute for Materials and Energy Sciences, SLAC National Accelerator Laboratory, 2575 Sand Hill Road, Menlo Park, CA 94025.}
\date{\today}

\begin{abstract}
The recent discovery of superconductivity in oxygen-reduced monovalent nickelates has raised a new platform for the study of unconventional superconductivity, with similarities and differences with the cuprate high temperature superconductors. In this paper we investigate the family of infinite-layer nickelates $R$NiO$_2$ with rare-earth $R$ spanning across the lanthanide series, introducing a new and non-trivial ``knob" with which to tune nickelate superconductivity. When traversing from La to Lu, the out-of-plane lattice constant decreases dramatically with an accompanying increase of Ni $ d_{x^2-y^2}$ bandwidth; however, surprisingly, the role of oxygen charge transfer diminishes. In contrast, the magnetic exchange grows across the lanthanides which may be favorable to superconductivity. Moreover, compensation effects from the itinerant $5d$ electrons present a closer analogy to Kondo lattices, indicating a stronger interplay between charge transfer, bandwidth renormalization, compensation, and magnetic exchange. We also obtain the microscopic Hamiltonian using Wannier downfolding technique, which will provide the starting point for further many-body theoretical studies.
\end{abstract}

\maketitle

\section{Introduction}

In 1986, the discovery of high-temperature superconductivity in the copper oxide (cuprate) La$_{2-x}$Sr$_x$CuO$_4$ (LSCO) by Bednorz and M\"uller ushered in a new era in condensed matter research \cite{BednorzMuller1986}.  Soon there after, a number of ``families'' of cuprates were discovered with electronic and structural similarities. While no consensus yet exists on the mechanism of superconductivity in these compounds, a number of characteristics seem to be common across the different families.  Structurally, they consist of quasi-two-dimensional CuO$_2$-planes separated by spacer, charge reservoir layers, which may contain a number of different atoms, including both rare-earth and/or oxygen.  Electronically, the low energy degrees of freedom primarily reside in the CuO$_2$-planes, where formal valence counting would yield Cu$^{2+}$ in a $3d^9$ electronic configuration; and crystal/ligand-field effects would place the orbitals with $d_{x^2-y^2}$ symmetry near the Fermi energy.  Characterized as charge transfer insulators within the Zaanen-Sawatzky-Allen (ZSA) scheme \cite{ZSA1985}, the undoped parent compounds are antiferromagnetic insulators due to strong correlations; and upon doping, one, or more, bands emerge and cross the Fermi level, depending on material specifics.  Aside from superconductivity, hole-doping in the cuprates produces a rich, complex phase diagram with a number of distinct, and potentially intertwined, phases \cite{Keimer2015}.

Research into unconventional superconductivity -- superconductivity that doesn't seem to fall within the BCS paradigm \cite{BCS}, as in the cuprates -- has evolved over time, and the list of materials now includes heavy fermion intermetallics, organic superconductors, ruthenates, and iron pnictides and chalcogenides.  Although the cuprates share some characteristics with these compounds, especially the proximity of the superconducting state to antiferromagnetism \cite{Scalapino_2012}, the lack of a material analog with both crystal and, more importantly, electronic structure similarities makes it more difficult to draw any general conclusions about a universal mechanism for unconventional superconductivity, if one exists at all.

Nickel oxide compounds, or nickelates, may provide such a platform, but it requires some unconventional chemistry.  In variants such as cubic NiO or single-layer La$_2$NiO$_4$, which is isostructural to the parent compound of the cuprate LSCO, the Ni-cation has a formal valence of Ni$^{2+}$ in a nominal $3d^8$ electronic configuration with spin $S=1$.  These nickelates can be doped as the cuprates, for instance with Sr or O, but these remain insulating. Guided by theoretical predictions\cite{Zaanen_a,Zaanen_b} this lead to the original discovery of the electronic stripes, turning out to be of the ``filled" insulating type \cite{Tranquada_a}. This in turn inspired the discovery of the similar ``spin stripes"  in the LSCO family \cite{Tranquada_b} with their intriguing relations to charge order in other cuprate families \cite{Comin}. A key difference was early on identified \cite{Zaanen_b} in the form of much stronger electron-phonon interaction associated with the holes in the nickelates as compared to the cuprates, rooted in the strongly Ni-O hybridized lower Hubbard band \cite{Zaanen_c}, leading to much stronger static lattice deformations stabilizing the filled stripes \cite{Zaanen_b,Tranquada_a}. A substantial literature emerged dedicated to the physics of these nickelate stripes (see e.g. the very recent study \cite{Zhang_2019}).  Recent reports even suggest that a three-layer variant possesses similar stripe-ordered ground states \cite{Zhang_2019}. 

Altering the chemical composition, LaNiO$_3$ is a rare-earth perovskite where the Ni-cation would have a formal valence of Ni$^{3+}$ in a $3d^7$ electronic configuration.  However, this electronic configuration would be energetically unfavorable, such that this compound is instead a negative charge transfer material with a nominal electronic configuration of $3d^8\ligand$, where $\ligand$ stands for a ligand-hole on oxygen \cite{Bisogni}. Unfortunately, such materials also show no signs of superconductivity. 

\begin{figure}
\includegraphics[width=\columnwidth]{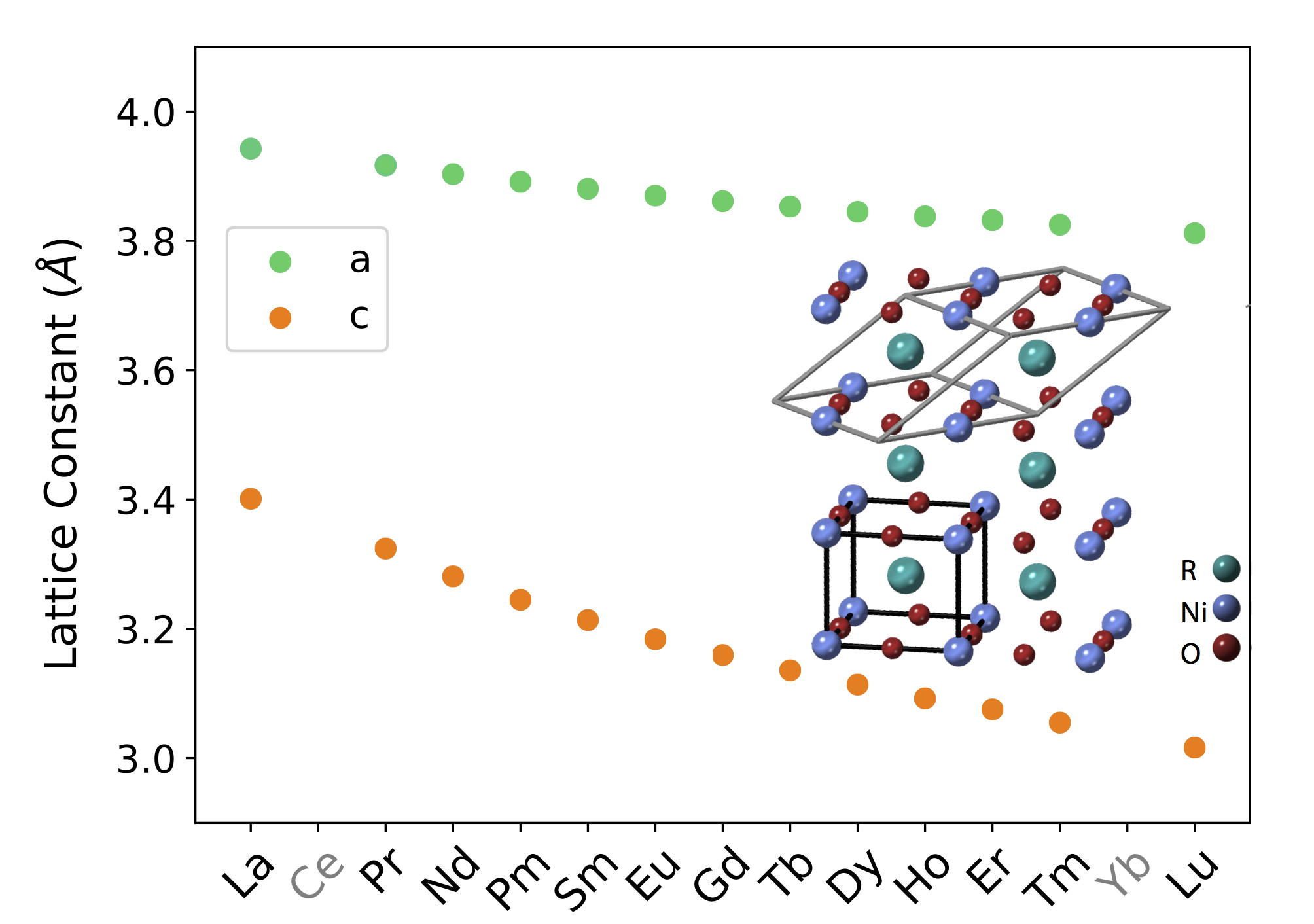}
\caption{\label{fig:figure1}  
{\bf{Lattice constant and crystal structure of RNiO$_2$.}} The DFT(GGA) relaxed RNiO$_2$ lattice constants. (inset) crystal structure for inifinite-layer nickelates RNiO$_2$. One Ni unit cell (black) for paramagnetic calculation and two Ni unit cell (gray) for ($\pi$,$\pi$,$\pi$) antiferromagnetic calculation are shown respectively. }
\end{figure}

Using a chemical reduction process \cite{ChemicalReduction1, ChemicalReduction2}, the apical oxygens can be removed from the rare-earth perovskite nickelates yielding rare-earth infinite layer nickelates {\R}NiO$_2$, where {\R} can be any of a number of rare-earth atoms.  These materials are isostructural to the infinite layer cuprates with quasi-two-dimensional NiO$_2$ planes separated by single layers of rare-earth atoms (see Fig.~\ref{fig:figure1}).  Formal valence counting yields Ni$^{1+}$ cations with a $3d^9$ electronic configuration similar to the Cu$^{2+}$ cations in the cuprates.  The formal similarities with the cuprates don't stop there, as the recent discovery of superconductivity in (Nd,Sr)NiO$_2$, with a $T_c$ as high as $15 \textrm{K}$ \cite{Danfeng2019}, potentially makes these infinite layer nickelates a new platform for studying unconventional superconductivity. 

Previous theoretical work, based on density functional theory (DFT) \cite{Rice, Pickett}, and recent experiments \cite{Hepting2019} have highlighted similarities between the {\R}NiO$_2$ compounds, primarily with {\R} = La or Nd, and infinite layer cuprates, as well as significant differences in their electronic structure.  A host of new theoretical proposals \cite{Norman, Raghu, Sawatzky, Wu, Arita, FuChunZhang, Vishwanath, Hoshino, Cano, ChenHanghui, DingYang, Held}, also primarily based on DFT, likewise find both similarities and differences to the cuprates.  While there is no overall consensus on the physics at this stage, there is more or less general agreement on a few observations about the electronic structure in {\R}NiO$_2$: in the absence of strong interactions, the low energy physics in the NiO$_2$ layers is that of a single, hole-like band crossing the Fermi energy; in the presence of strong interactions, the NiO$_2$ layers become ``Mott'' insulating, although such a state still involves a significant amount of oxygen character \cite{Norman}; and unlike the cuprates, rather than sitting well above the Fermi energy, the rare-earth band forms small, but significant, metallic pocket(s) \cite{Pickett, Hepting2019}.  Of course, this leaves much to explore.  In particular, lacking any experimental signatures of magnetic order in the parent compounds \cite{ChemicalReduction1, HAYWARD2003839}. what is the nature of magnetism, if any, in {\R}NiO$_2$ and how might it evolve with changes in chemical composition and doping?  What role does the metallic rare-earth band play, given it's small size, and what about quasiparticle compensation and extinction?  Ultimately, can one infer something universal about unconventional superconductivity from the rare-earth nickelates? Answering these questions, and many others, will require a great deal more experimental, and theoretical, work. 

To at least begin probing for clues to the underlying physics in the rare-earth nickelates, and lifting the veil on general principles of unconventional superconductivity, here, we present a systematic investigation into the parent compounds {\R}NiO$_2$, where {\R} runs across the entire lanthanide series. Based on our electronic structure calculations, we show that two bands near $E_F$ contribute to the physics of these materials. One quasi-two-dimensional band, comprised of Ni $3d_{x^2-y^2}$ and in-plane O $2p_x$ and $2p_y$ orbitals, has a Zhang-Rice-singlet-like character, but with a smaller oxygen contribution than in the cuprates, which decreases going from La to Lu. The bandwidth of this band increases, as does the in-plane magnetic exchange, also going from La to Lu.  The other band, comprised of rare-earth $5d$ orbitals, crosses $E_F$ forming a small, but significant metallic electron pocket. This $5d$ band has a fully three-dimensional dispersion, corresponding to an itinerant electron picture with three dimensional long range hoppings. These two bands hybridize, possibly forming a new type of Kondo- or Anderson-lattice system, where the strongly correlated NiO$_2$ layer plays the role of the $4f$-electrons in the heavy fermion, rare-earth intermetallics.   The electron density in the itinerant band is quite low, so it is unclear what effect it may have on the strongly correlated Ni $3d_{x^2-y^2}$ band, which is more reminiscent of the low energy electronic structure in the cuprates.  Finally, we present a microscopic Hamiltonian, and effective parameters for representative compounds, which can serve as a starting point for more complex many-body calculations for specific materials and the infinite layer rare-earth nickelate family, in general.

\section{Crystal Structure}

The parent compounds of the infinite layer rare-earth nickelates {\R}NiO$_2$ have four atoms in a primitive unit cell with space group Pmmm:  Ni at (0,0,0); O at ($\frac{1}{2}$,0,0) and (0,$\frac{1}{2}$,0); and {\R} at ($\frac{1}{2}$,$\frac{1}{2}$,$\frac{1}{2}$); all in units of the lattice constants $a$, $b$, and $c$, respectively, with $a$ = $b$, as shown schematically in the inset of Fig.~\ref{fig:figure1}.  One expects that the size of the rare-earth atom decreases with increasing atomic mass $Z$.  We optimized the lattice constants for compounds across the lanthanide series using density functional theory (DFT) as implemented in Quantum ESPRESSO \cite{QE}, with PAW pseudopotentials. As we will discuss in more detail, we choose to place the 
$4f$ electrons in the core.  As shown in Fig.~\ref{fig:figure1} this optimization yields a systematic reduction of the out-of-plane lattice constant $c$ by $\sim 10\%$ across the lanthanide series, as well as a reduction of the in-plane lattice constant $a$ by $\sim 5\%$, assuming bulk material.  Thin films grown on an SrTiO$_3$ substrate (experimental in-plane lattice constant $\sim 3.91 \textrm{\AA{}}$ and GGA calculated in-plane lattice constant $\sim 1\%$ larger \cite{GGASTO}) for most of the lanthanide series would experience an in-plane tensile strain, which would tend to further reduce the out-of-plane lattice constant across the series.  

\begin{figure*}

\begin{overpic}[width=1.9\columnwidth]{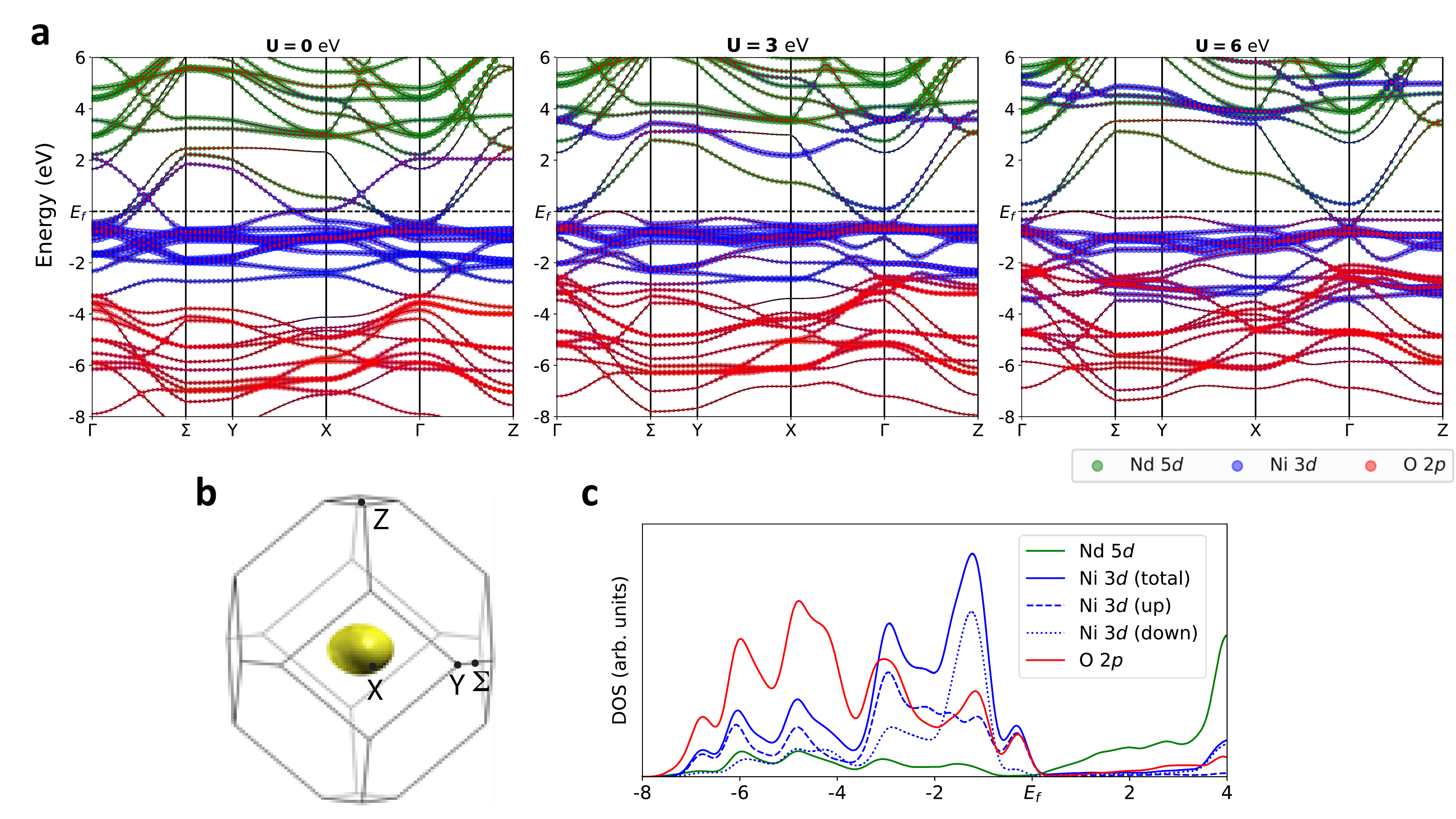}
    \put(140,-30){\includegraphics[width=0.5\columnwidth, trim={1.5cm 0 8cm 3.7cm},clip]{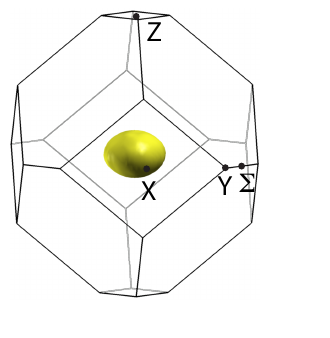}}
\end{overpic}

\caption{\label{fig:figure2}  
{\bf{Electronic structure of NdNiO$_2$.}} a, GGA and GGA+$U$($U$= 3 and 6 eV) calculations of bandstructure and the corresponding atomic orbital character of NdNiO$_2$.  The ``lower Hubbard band" and ``upper Hubbard band" splitting in DFT+U scenarios are apparent, specifically for $U$=6 eV they are located at $\sim$ -0.5 - 0 eV and $\sim$ 3.5 - 5 eV relative to $E_f$ respectively. b, Visualization of Fermi surface geometry in the first BCT BZ for NdNiO$_2$ in GGA+$U$ (6 eV) calculation. c, Partial density of states of NdNiO$_2$ in GGA+$U$ (6 eV) calculation, included are contributions from spin-polarized Ni 3\textit{d} states on one Ni atom to elucidate the symmetry in the two-Ni cell. Summary, the orbital character of the bands at Fermi surface is dominant by Nd 5$d$, forming the electron pocket at the zone center.
It is notable that this electron pocket becomes somewhat smaller but does not disappear with a larger value of $U$. In contrast, the nickel 3\textit{d} bands just below the Fermi energy hardly change with different $U$.}
\end{figure*}

Given the stacking, one expects that the out-of-plane lattice constant should reflect decreasing radius of the trivalent  rare-earth ions going to the right in the series. This actually matches the  $\simeq$ 10\% contraction of the lattice constants of the elemental rare-earth metals quite well \cite{Spedding}. However, this is in striking contrast with the {\R}Ba$_2$Cu$_3$O$_{7-\delta}$ system where the changes in lattice constants are on the 1 percent level -- accordingly, it was early on established that rare-earth substitution makes very little difference for the physical properties (including $T_c$'s) in this cuprate family \cite{PHHor}.  One may suspect that this substantial change of volume going through the series may have a more substantial relationship with the overall electronic structure in comparison to the $123$ cuprates. 
This will indeed turn out to be the case according to our computations.

\section{Electronic Structure}

\subsection{General Electronic Structure Features from DFT+U Calculations}

The electronic structure of {\R}NiO$_2$ has been the subject of study for some time \cite{Rice, Pickett}.  Generally, it has been recognized that the electronic structure near the Fermi energy consists of several features, including Ni $3d$, O $2p$ and {\R} $5d$ \cite{Hepting2019, Pickett}. To investigate the evolution of these features with changes in the rare-earth element, and properly account for correlation effects in the quasi-two-dimensional NiO$_2$ plane, we performed density functional mean-field calculations, including a Ni $3d$ on-site Hubbard $U$, using the LDA+$U$ technique \cite{DFTUZaanen}, which can provide information on general trends in the electronic structure across the lanthanide series.  

\begin{figure*}
\includegraphics[width=2\columnwidth]{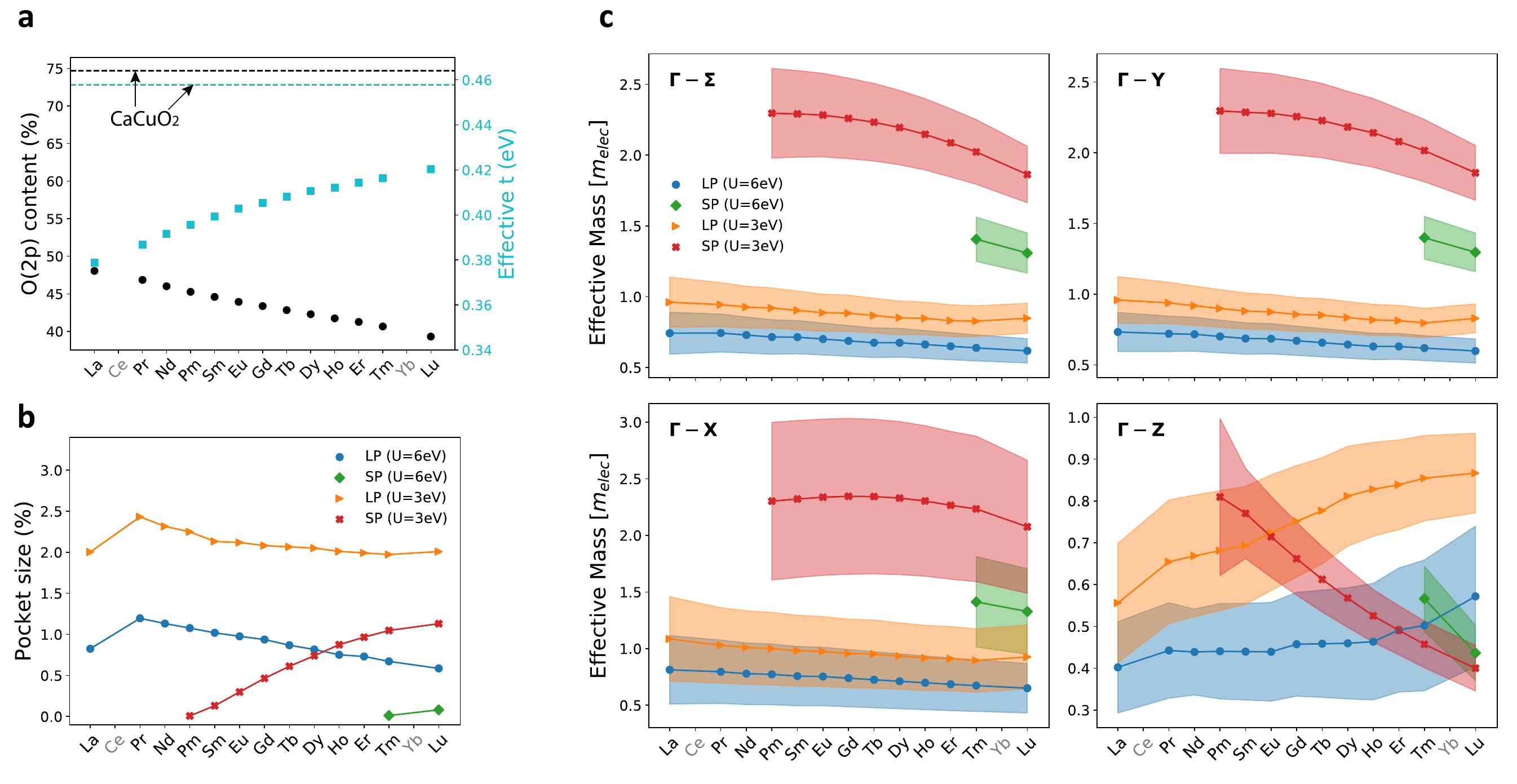}
\caption{\label{fig:figure3}  
{\bf{Electronic structure of {\R}NiO$_2$.}} a, (left axis) The oxygen 2$p$ content in percentage of the total for the lower Hubbard model in the antiferromagnetic GGA+$U$ ($U$ = 6eV) calculation for {\R}NiO$_2$ and CaCuO$_2$. (right axis) Effective Ni $d_{x^2-y^2}$ nearest-neighboring hopping as estimated by $\frac{1}{8}\times$bandwidth of the Ni $d_{x^2-y^2}$ band in the paramagnetic GGA calculation. Effective Cu $d_{x^2-y^2}$ nearest-neighboring hopping as estimated by $\frac{1}{8}\times$bandwidth of the Cu $d_{x^2-y^2}$ band in the paramagnetic GGA calculation is shown in the dashed line. b, Rare-earth pocket size in percentage of the first BZ volume. Large pocket (LP) is present for all elements in the lanthanide series for both $U$ = 3eV and 6eV. Small pocket (SP) appears at Pm and the elements to the right for $U$ = 3eV. SP appears at Tm and the elements to the right for $U$ = 6eV. (c) Rare-earth electron pocket effective mass in all principal axes directions. See Methods section for more information about how data processing was conducted.}
\end{figure*}

Fig.~\ref{fig:figure2}a illustrates the electronic structure of NdNiO$_2$ along high-symmetry lines within the body centered tetragonal (BCT) first Brillouin zone (BZ) for $U$ = 0, 3 and 6 eV, respectively. The orbital character of the bands is superimposed on the electronic structure. Similar sequences for the other infinite layer rare-earth nickelates, but without orbital character, appear in the Appendix, Fig.~\ref{fig:app_figure1}.  Although no explicit folding occurs for $U$ = 0 eV, we plot the bandstructure in the two-Ni unit cell for ease of comparison.  Starting with $U$ = 0 eV, one observes four bands crossing the Fermi level:  (1) two three-dimensional ``Nd-bands'', forming pockets near the $\Gamma$-point; and (2) two quasi-two-dimensional ``Ni-bands''.  The two three-dimensional pockets with predominantly Nd-character form from original pockets at the $\Gamma$-point and A-point in the unfolded tetragonal BZ.  Note that only the Nd-bands cross the Fermi energy along the $\Gamma$-Z cut of Fig.~\ref{fig:figure2}a. The two ``Ni-bands'' have mixed Ni and O-character.  We will return to this point later.  Note a significant difference in the van Hove energy at the X-point between the original and folded band ($\sim$ a few hundred meV), highlighting the quasi-two-dimensional nature of these bands.

While the precise value for the Hubbard $U$ is not established, one can make reasoned guesses.  The value likely would be larger than that of Ni metal ($\sim$ 2 eV), due to better screening of the interaction in the metal compared to {\R}NiO$_2$ compounds.  It likewise would be smaller than the value in NiO ($\sim$ 8 eV), a prototypical charge transfer insulator in the ZSA scheme \cite{ZSA1985}.  Fig.~\ref{fig:figure2}a shows the bandstructure for NdNiO$_2$ while systematically increasing the Hubbard $U$.  First, the strong Coulomb repulsion forces the Ni-bands to form characteristic upper and lower ``Hubbard bands'' separated by $\sim U$.  As with the O-character, we will return to the ZSA classification later and simply refer to these as upper and lower Hubbard bands.  These bands progressively renormalize and flatten with increasing Hubbard $U$.  At the same time, the Nd-pockets become progressively smaller; and one disappears entirely for $U$ = 6 eV, with the band bottom just above the Fermi level at the $\Gamma$-point.  The Fermi surface for $U$ = 6 eV consists of a single, electron-pocket centered at the $\Gamma$-point with predominantly Nd-character as shown in Fig.~\ref{fig:figure2}b.

Fig.~\ref{fig:figure2}c shows the orbitally resolved partial density of states with $U$ = 6 eV.  First, note the small (albeit significant) Nd density of states at the Fermi energy, corresponding to the small Nd-pocket in Fig.~\ref{fig:figure2}d.  Concentrating on Ni and O, the O $2p$-bands lie much further away from the Fermi energy than in other nickelates \cite{Hepting2019, LaNiO3}, or more importantly than the cuprates \cite{Emory}, marking a system with significantly decreased oxidation level. This is further confirmed by the mixing between Ni and O in the lower Hubbard band is much smaller compared to prototypical charge transfer insulators like the cuprates. Specifically, the O content in the lower Hubbard band is much smaller in {\R}NiO$_2$ (between 40\% and 48\%) than in CaCuO$_2$ ($\sim$ 75\%) according to our DFT+U calculations, as shown in Fig. 3a. These observations place NdNiO$_2$ (and more generally {\R}NiO$_2$)  well inside the Mott-Hubbard regime of the ZSA classification. As highlighted recently \cite{Norman}, as in the cuprates one may view
the doped holes as Zhang-Rice singlet-like states which are however now better defined because of their smaller O-like character.  The results presented here are very much in-line with such a picture, showing reduced O-character in the lower Hubbard band, as well as recent Ni $L$-edge x-ray absorption (XAS) measurements, which confirm the $3d^9$-like state of Ni, and O $K$-edge XAS, which highlights significant differences between the O-character in infinite layer nickelates when compared to other charge transfer compounds, and the cuprates \cite{Hepting2019}.

\subsection{Trends across Lanthanide Series: Bandwidth, Oxygen Content, and R Pocket}

A systematic survey across the lanthanides of the bandstructure with $U$ = 0, 3 and 6 eV and the density of states for the $U$ = 6 eV case is presented in the Appendix, Figs.~\ref{fig:app_figure1}.  While variations exist with changes in the rare-earth element, Fig.~\ref{fig:figure3} highlights some of the most important trends.  For $U$ = 0 eV (``non-interacting'' Ni-bands), one can extract the effective bandwidth for the quasi-two-dimensional Ni-bands, which includes not only the in-plane dispersion, but also the contributions from any out-of-plane hybridization.  The trend appears in Fig.~\ref{fig:figure3}a (left-axis).  While the bandwidth is $\sim$ 3 eV for all compounds, it monotonically increases by $\sim$ 10\% across the lanthanides.  This should come as no surprise considering the trend in lattice constants shown in Fig.~\ref{fig:figure1}:  a reduction in both the in-plane, and out-of-plane, lattice constants should lead to increased orbital overlap, and by extension, hybridization and bandwidth.

Let us now turn to the big surprise revealed by this band structure study. Given that the lattice is contracting while the bandwidth is increasing going toward the end of the series,
one would expect that the degree of covalency associated with the holes in the Ni-O planes should also increase. This is however not the case at all. The degree of oxygen admixture is easy to quantify from the band structure by inspecting the orbital content of the lower Hubbard band and the result is shown in Fig.~\ref{fig:figure3}a. This oxygen 
character is very large to the left of the series, as high as $\sim$ 48\% in La. 
But we observe that it  decreases by as much as $\sim$ 40\% running across the lanthanide series. One expects such differences when one goes down in the periodic system substituting e.g. O for S, but not with the substitution by different rare earth, famous for their chemical similarity. 
 Fig.~\ref{fig:app_figure1} in Appendix may provide a clue regarding this apparent paradox.  Observing the distribution of the O weight on a large energy scale one can discern a trend:
 running across the lanthanide series, the centroid of O weight monotonically moves to higher binding energy, by perhaps 1 eV from La to Lu.  Such a change would lead to an increase in the charge transfer energy and would offset any increases in hybridization relating to the mixing between Ni and O. Given the unusual chemistry of these mono-valent nickelates the chemically ``inert" lanthanides appear to exert a spectacular influence on the nature of the oxidation in the Ni-O perovskite planes.  

Another surprise associated with the lanthanide  substitution revolves around the evolution of the {\R} 5$d$ electron pockets. We already highlighted that for La these pockets are sensitive to $U$. For $U$= 0 eV, GGA indicates the presence  of ``large" volume (LP) and ``small" volume (SP) pockets. However, for $U$  = 6 eV the small pocket has disappeared and only the LP remains, as indicated in Fig.~\ref{fig:figure2}b. In Figs.~\ref{fig:figure3}b and c we characterize what happens with these pockets going through the series for $U$ = 3 and 6 eV. Focusing on the volume enclosed by the LP (orange and blue) we see from Fig.~\ref{fig:figure3}b  that its volume is decreasing for increasing $U$, but the dependence of this volume on the {\R} is weak. The surprise is that upon moving through the series there is a critical point where the small pocket re-emerges. This happens for $U=3$eV at Pm while for $U=6$eV one has to go all the way to the end of the series.   Finally, we analyze the effective mass of these pockets at $k_{\textrm{F}}$ along the high-symmetry directions in the BCT BZ  (Fig.~\ref{fig:figure3}c). We
find that in the in-plane directions not much happens upon changing the {\R}. All of the action is in the $c$ direction: the mass associated with the LP shows a significant increase in the series, while the SP shows a quite strong decrease upon going towards the end of the series.

Notably, these changes to the electronic structure cannot be explained by volume change alone. When the rare-earth $R$ is kept constant and the lattice parameters are changed by an amount comparable to those shown in Figure \ref{fig:figure1}, the second electron pocket never appears, small hole pockets start to appear along $\Gamma-\Sigma$ and $Y-X$, and the effective hopping $t$ is significantly lower.

\begin{figure}[h]
\includegraphics[width=\columnwidth]{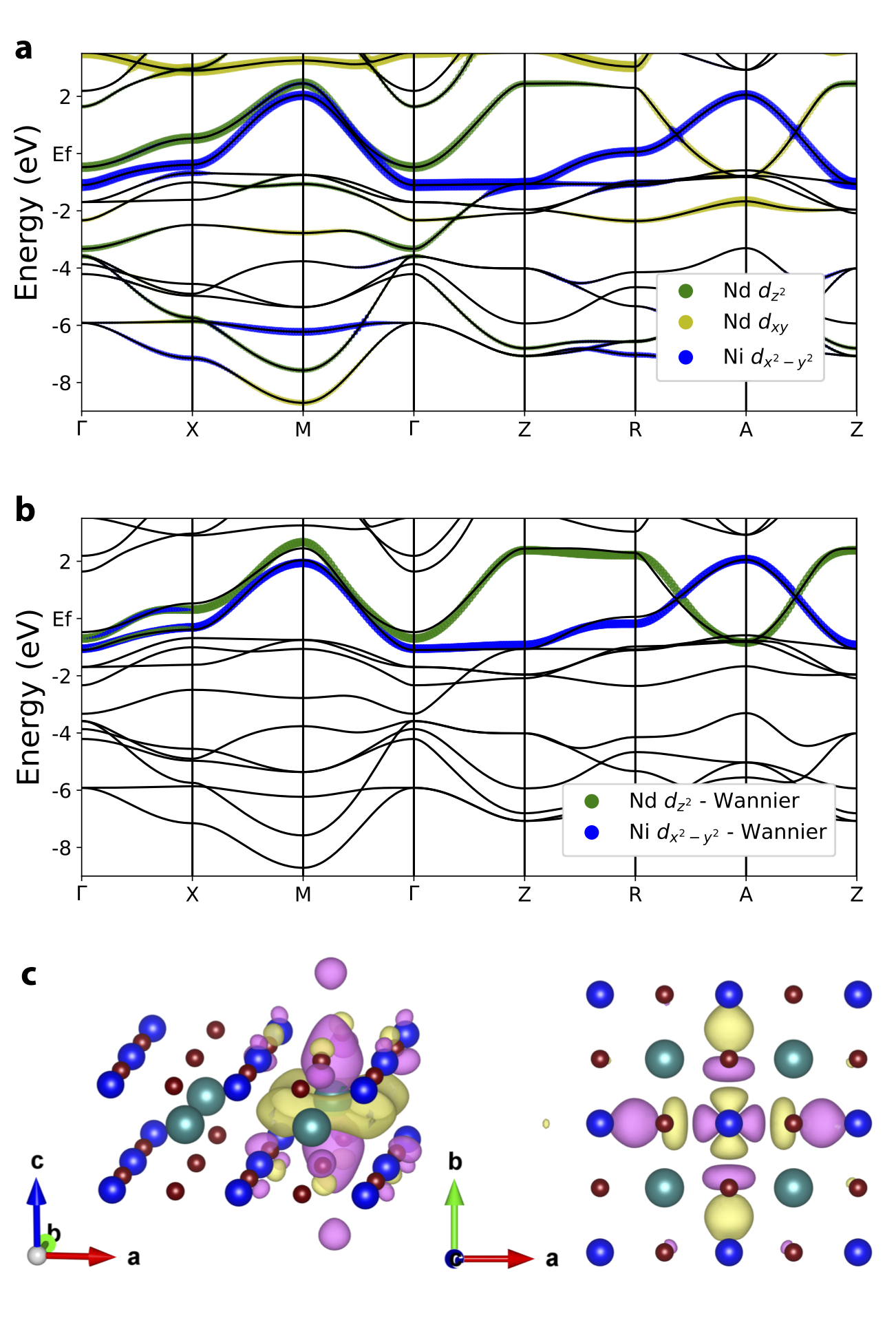}
\caption{\label{fig:figure4}  
{\bf{Wannier downfolding of NdNiO$_2$.}} a, GGA calculation of NdNiO$_2$ band bandstructure, and Nd $d_{z^2}$ (green), Nd $d_{xy}$ (yellow) and Ni $d_{x^2-y^2}$ (blue) atomic orbital content. b, The dispersion of two-orbital model based on the Wannier downfolding on the Wannier orbitals shown in c. c, Nd $d_{z^2}$ -like and Ni $d_{x^2-y^2}$ -like Wannier orbitals.}
\end{figure}

\section{Microscopic Hamiltonian}

Based on the electronic structure discussed above, we propose a model for {\R}NiO$_2$, involving two-orbitals with an on-site Hubbard $U$ associated with the planar NiO$_2$ band (Ni-band):

\begin{eqnarray}
H &=& \sum_{k,\sigma} (\varepsilon^{\R}_k n^{\R}_{k,\sigma} + \varepsilon^{Ni}_k n^{Ni}_{k,\sigma}) + U\sum_i n^{Ni}_{i,\uparrow}n^{Ni}_{i,\downarrow} \nonumber\\
&& + \sum_{k,i,\sigma} (V_{k,i}c^\dagger_{k,\sigma}d_{i,\sigma} + h.c.),\label{eq:equation1}
\end{eqnarray}
where the first term describes the non-interacting {\R}- and Ni-bands, the second term represents the on-site Hubbard interaction in the Ni-band, and the last term describes the hybridization between the Ni- and {\R}-bands.  Here, $\varepsilon^{\alpha}_{k}$ represents the momentum resolved ``non-interacting'' bandstructure, $n^{\alpha}_{k,\sigma}$ is the usual number operator, and the hybridization term has been written to resemble the typical form for the Kondo- or Anderson-lattice model with a hybridization between the dispersive {\R} $5d$-band, with $c_k$ ($c^{\dagger}_k$) operators and the Hubbard-like Ni $3d$-band, with $d_i$ ($d^{\dagger}_i$) operators.

We derive parameters for the microscopic Hamiltonian by performing Wannier downfolding on the bandstructure in the one Ni unit cell with $U$ = 0 eV. Fig.~\ref{fig:figure4}a shows the bandstructure and the atomic orbital content of NdNiO$_2$ from DFT calculations. The Ni $d_{x^2-y^2}$ band highlighted in blue crosses $E_F$. The other band crossing $E_F$ has mostly Nd $d_{3z^2-r^2}$ character with some Nd $d_{xy}$ component close to the A-point. Wannier downfolding for the two bands was then performed to obtain the microscopic Hamiltonian on the two maximally-localized Wannier orbitals as implemented in Wannier90 \cite{Wannier90}, using atomic Ni $d_{x^2-y^2}$ and Nd $d_{3z^2-r^2}$ as the initial configuration. The Ni $d_{3z^2-r^2}$ band was not included because its on-site energy is much lower than Ni $d_{x^2-y^2}$ \cite{Hepting2019} and, although Ni $d_{3z^2-r^2}$ weight appears near E$_{f}$, this comes primarily from the strong hybridization between Ni $d_{z^2}$ and $R$ $d_{z^2}$. This weight will be incorporated naturally in the Wannier orbital centered on $R$ $d_{z^2}$ when downfolded to a two-orbital model. Further symmetrization to reserve the $D_{4h}$ symmetry for the Nd $d_{3z^2-r^2}$ Wannier orbital and the corresponding Hamiltonian parameters are performed. The converged Wannier downfolding yields the dispersion as shown in Fig.~\ref{fig:figure4}b, with green highlighting the Nd-character and blue highlighting the Ni-character. The two bands mix significantly along $\Gamma$-$X$. The corresponding Wannier orbitals are shown in Fig.~\ref{fig:figure4}c, where by construction the Nd Wannier orbital has $d_{3z^2-r^2}$ symmetry centered on Nd, and extends out-of-plane, with a rounded square shape in-plane, while the Ni Wannier orbital has $d_{x^2-y^2}$ symmetry centered on Ni, and extends onto the neighboring oxygens with $2p$ symmetry. We have performed the Wannier downfolding approach on other {\R}NiO$_2$ materials for {\R} = La, Pr, Eu, and Dy selectively. The parameters for these {\R}NiO$_2$ are shown in Table~\ref{tab:table1}. The table lists values above a minimum threshold of 10 meV, such that the non-interacting {\R}- and Ni-band dispersions take the form:

\begin{table*}
\begin{widetext}
\begin{eqnarray}
\varepsilon^{R}_{k}&=&\varepsilon^R_{0}
+ 2\, t^{R}_{[0,0,1]}\,\cos(k_{z})
+ 2\, t^{R}_{[0,0,2]}\,\cos(2\,k_{z})
+ 2\, t^{R}_{[0,0,3]}\,\cos(3\,k_{z}) \nonumber \\
&\quad+& \big[2\, t^{R}_{[1,0,0]}
+ 4\, t^{R}_{[1,0,1]}\,\cos(k_{z})
+ 4\, t^{R}_{[1,0,2]}\,\cos(2\,k_{z})\big] \big[\cos(k_{x}) + \cos(k_{y})\big] \nonumber \\
&\quad+& \big[4\, t^{R}_{[1,1,0]}
+ 8\, t^{R}_{[1,1,1]}\,\cos(k_{z})
+ 8\, t^{R}_{[1,1,2]}\,\cos(2\,k_{z})
+ 8\, t^{R}_{[1,1,3]}\,\cos(3\,k_{z})\big] \cos(k_{x})\,\cos(k_{y}) \nonumber \\
&\quad+& \big[2\, t^{R}_{[2,0,0]}
+ 4\, t^{R}_{[2,0,1]}\,\cos(k_{z})\big] \big[\cos(2\,k_{x}) + \cos(2\,k_{y})\big] \nonumber \\
&\quad+& \big[4\, t^{R}_{[2,1,0]}
+ 8\, t^{R}_{[2,1,1]}\,\cos(k_{z})\big] \big[\cos(2\,k_{x})\,\cos(k_{y}) + \cos(k_{x})\,\cos(2\,k_{y})\big]
\end{eqnarray}

\begin{eqnarray}
\varepsilon^{Ni}_{k}&=&\varepsilon^{Ni}_{0} + 2\, t^{Ni}_{[1,0,0]}\,\big[\cos(k_{x}) + \cos(k_{y})\big]
+ 4\, t^{Ni}_{[1,1,0]}\,\cos(k_{x})\,\cos(k_{y})
+ 2\, t^{Ni}_{[2,0,0]}\,\big[\cos(2\,k_{x}) + \cos(2\,k_{y})\big] \nonumber \\
&\quad+& \big[2\, t^{Ni}_{[0,0,1]}
+ 4\, t^{Ni}_{[1,0,1]}\,\big[\cos(k_{x}) + \cos(k_{y})\big]
+ 8\, t^{Ni}_{[1,1,1]}\,\cos(k_{x})\,\cos(k_{y})\big] \cos(k_{z})
\end{eqnarray}
\end{widetext}

 \centering
\begin{tabular}{| p{1cm} | p{1cm} | p{1cm} | p{1.5cm} | p{1.5cm} | p{1.5cm} | p{1.5cm} | p{1.5cm} |}
\hline
\multicolumn{8}{| c |}{\rule{0pt}{3ex}Hopping Parameters for Wannier Downfolding} \\
\hline
\multicolumn{3}{| c |}{\rule{0pt}{3ex}} & La & Pr & Nd & Eu & Dy \\
\hline
\hline
i & j & k & \multicolumn{5}{| c |}{\rule{0pt}{3ex}$t^{R}_{[i,j,k]}$ (eV)} \\
\hline
\rule{0pt}{3ex}0 & 0 & 0 & 1.132  & 1.280   & 1.305  & 1.360  & 1.413\\
0 & 0 & 1 & -0.022 & -0.132 & -0.164 & -0.214  & -0.240\\
0 & 0 & 2 & -0.112 & -0.153 & -0.172  & -0.185 & -0.223\\
0 & 0 & 3 & 0.019  &            &             &             &  \\
1 & 0 & 0 & -0.028 & -0.025 & -0.028 & -0.033  & -0.016\\
1 & 0 & 1 & -0.164 & -0.214 & -0.206 & -0.209  & -0.220\\
1 & 0 & 2 & 0.033  & 0.035  & 0.030   & 0.036   & 0.043\\
1 & 1 & 0 & -0.062 & -0.079 & -0.090 & -0.061  & -0.082\\
1 & 1 & 1 & 0.024  & 0.060  & 0.066  & 0.052   & 0.068\\
1 & 1 & 2 & 0.013  & 0.019  & 0.021  & 0.017    & 0.011\\
1 & 1 & 3 & -0.019 &            &            &              &  \\
2 & 0 & 0 &            & 0.016  & 0.027  & 0.023    & 0.029\\
2 & 0 & 1 & 0.009  & -0.007 &  -0.012  & -0.009  & -0.010\\
2 & 1 & 0 &            & -0.014 & -0.021 & -0.020  & -0.023\\
2 & 1 & 1 & 0.009  &            &             &             &  \\
\hline
\hline
i & j & k & \multicolumn{5}{| c |}{\rule{0pt}{3ex}$t^{Ni}_{[i,j,k]}$ (eV)} \\
\hline
\rule{0pt}{3ex}0 & 0 & 0 & 0.267  & 0.312 & 0.308    & 0.314  & 0.319 \\
1 & 0 & 0 & -0.355 & -0.376 & -0.374 & -0.370 & -0.385 \\
1 & 1 & 0 & 0.090  & 0.079 &  0.094   & 0.072  & 0.095 \\
2 & 0 & 0 & -0.043 & -0.033 & -0.043 & -0.036 & -0.041 \\
0 & 0 & 1 & -0.043 & -0.038 & -0.033 & -0.044 & -0.061 \\
1 & 1 & 1 & 0.013  & 0.021  &             &  0.026  &   \\
1 & 0 & 1 &            &            &             & -0.020 &  \\
\hline
\hline
i & j & k & \multicolumn{5}{| c |}{\rule{0pt}{3ex}$t^{R-Ni}_{[i,j,k]}$ (eV)} \\
\hline
\rule{0pt}{3ex}2 & 0 & 0 & -0.026 & 0.021 & 0.020   & 0.017 & 0.016\\
2 & 0 & 2 & 0.013 &            &              &          &  \\
\hline
\end{tabular}
\caption{\label{tab:table1} \textbf{Hopping parameters for two-orbital Wannier downfolding of $R$NiO$_2$ ($R$ = La, Pr, Nd, Eu and Dy)}. The hopping terms within Ni or {\R} bands represent the matrix elements as shown in Eq. (2) and (3). The cross terms represent the matrix elements for $\sum_{i_0, j_0, k_0, i,j,k}t^{R-Ni}_{[i,j,k]}  c^{\dagger R}_{i_0, j_0, k_0} c^{Ni}_{i_0+i, j_0+j, k_0+k}$ + symmetrically equivalent terms + h.c.. The fractional coordinates are {\R} = [0.5,0.5,0.5] and Ni = [0,0,0] in the paramagnetic unit cell.}
\end{table*}

Recently, different model Hamiltonians have been proposed for the low energy physics of infinite-layer nickelates \cite{Norman, Raghu, Sawatzky, Wu}, although our model Hamiltonian is functionally unique. Including Ni $d_{x^2-y^2}$ and {\R} $d_{z^2}$ is universal as they capture the main features of the bands across $E_F$, according to both paramagnetic DFT calculations in single Ni unit cells and antiferromagnetic DFT(+$U$) calculations in two Ni unit cells. Some theories take a multi-orbital model (more than two orbitals) including even Ni $d_{z^2}$ or {\R} $d_{xy}$ into consideration.  
We note that our two-orbital downfolded model can capture the contributions from {\R} $d_{xy}$ and Ni $d_{z^2}$ as well. Our {\R} $d_{z^2}$-like Wannier orbital has non-neglible contribution on the Ni $d_{z^2}$ character, as shown in Fig. 4c, naturally capturing this Ni $d_{z^2}$ contribution. Moreover, based on the DFT+$U$ bandstructure, the Ni $d_{z^2}$ contribution only appears at the bottom of the upper valence band at the X-point, making it less relevant in the case of light hole doping.  Our downfolded {\R} $d_{z^2}$-like Wannier orbital, as shown in Fig. 4c, has lower symmetry than the atomic $d_{z^2}$ orbital, capturing some contribution from $d_{xy}$ orbital. Our two-orbital model, as shown in Figure 4, also captures well the dispersion around the A-point, where the $R$ $d_{xy}$ pocket is present. For these reasons, we believe that our proposed two-orbital model is the simplest model capturing the correct orbital contribution for infinite-layer nickelates.

\section{Magnetism and Superexchange}

Regardless of one's specific beliefs, there is a general consensus that magnetism is a key aspect underlying the physics of the cuprates \cite{LaTecon2012, Tranquada1996, Aeppli1997, Kenzelmann}.  
As already mentioned, there are no experimental results yet shedding light on the magnetism of the infinite layer rare-earth nickelates \cite{Danfeng2019}, although this continues to be the subject of aggressive experimental inquiry. Nevertheless, our density functional theory calculations using LDA+$U$ possess, and our simple effective model likely may possess, an antiferromagnetic ground state.  Recent articles have attempted to address the subject of superexchange and magnetism, specifically for NdNiO$_2$ \cite{Raghu, Wu, Sawatzky}.  Some theories highlight similarities to the cuprates \cite{Raghu}. while others report substantive differences,
ranging from a non-monotonic doping dependence of the magnetism \cite{Pickett} to an order-of-magnitude reduction in the effective superexchange \cite{Sawatzky}.

Our effective model Eq.~\ref{eq:equation1} as based on first principles bares resemblances to a Kondo- or Anderson-lattice model. Well known from the heavy-fermion intermetallics,
 the Kondo-coupling  between the itinerant valence electrons and the localized $f$-electrons/spins may lead to magnetism and magnetic behavior driven by the RKKY interaction \cite{RKKY1, RKKY2, RKKY3}, in turn competing with the tendency to form non-magnetic Kondo-singlets \cite{Doniach}. Compared to these main-stream systems, 
 two factors in the electronic structure of the {\R}NiO$_2$ series set it apart:  (1) the relatively small size of the rare-earth pocket and (2) the single-band-Hubbard-like nature of the electronic structure in the NiO$_2$-layers.
 
The modest pocket size implies that the Kondo screening should be limited given the fact that there is only a fraction of an itinerant electron available per localized spin (Noziere's exhaustion principle \cite{Nozieres1985, Nozieres1998}). This ``Kondo physics" may be quite sensitive to the details of these pockets and one may anticipate that it will therefore be sensitive to the rare earth substitution (see Fig.~\ref{fig:figure3}b).
 
The other novelty is the large direct hybridization between the Ni $d^9$ states. This will lead to strong superexchange interactions between the Ni spins which we predict to be large
 compared to the Kondo-couplings involving the electron pockets. For this reason one may better view the situation as being closely related to the effective single-band Hubbard 
 models used to describe the low energy physics in the cuprates, enriched by corrections coming from the small pocket Kondo physics. Given the separation of scales, the consequences
 of these ``Kondo corrections" on the physics of (doped) Mott insulators should be tractable to quite a degree using perturbation theory, but this particular situation appears to be presently,
 completely unexplored. 
 
\begin{figure}[h]
\includegraphics[width=\columnwidth]{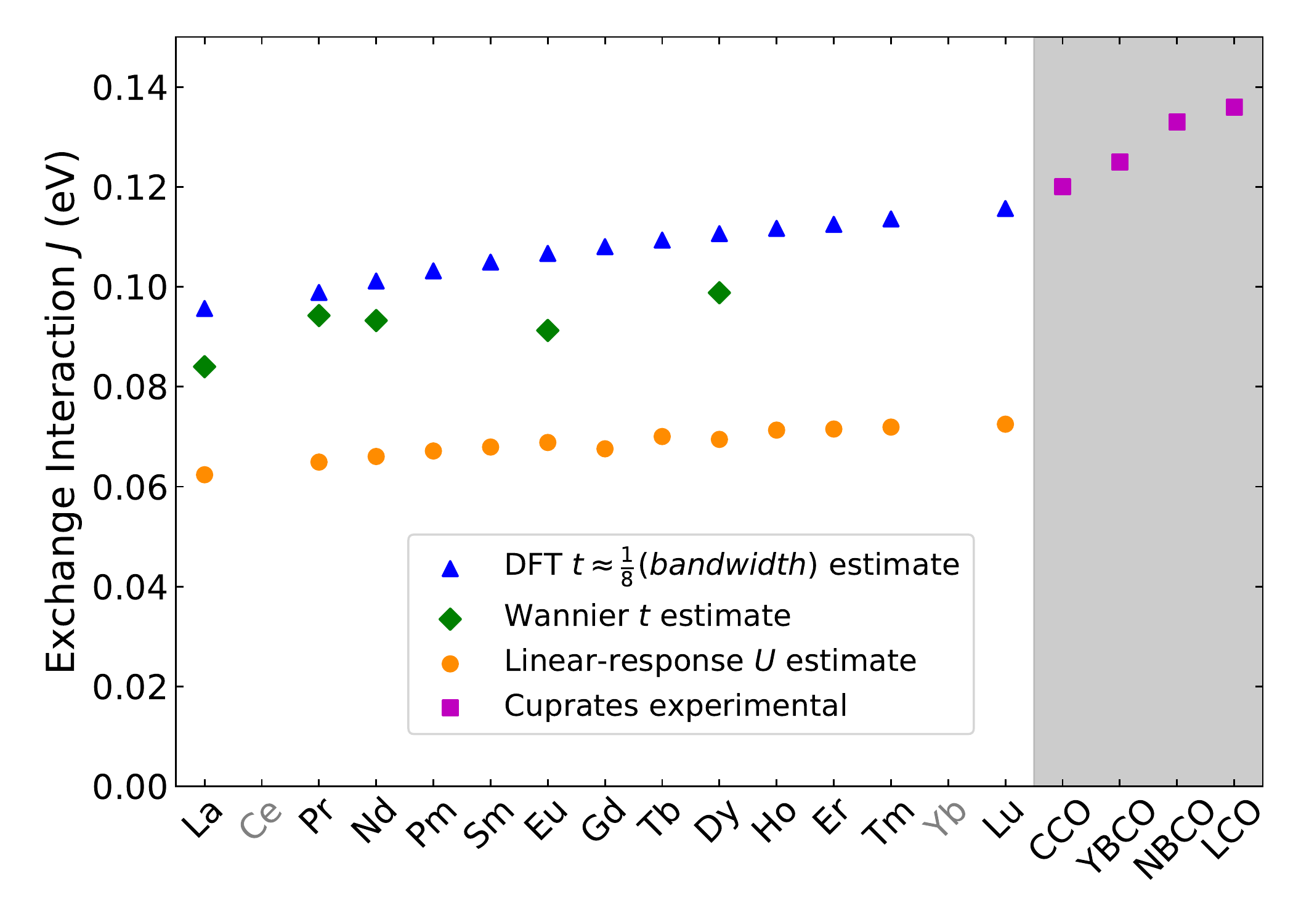}
\caption{\label{fig:figure5}  
{\bf{Exchange interaction of RNiO$_2$.}} (blue) Exchange interaction $J$ calculated by $4t^2/U$, where $t$ is estimated by $\frac{1}{8}\times$bandwidth of Ni 3$d_{x^2-y^2}$ band. (green) Exchange interaction $J$ calculated by $4t^2/U$, where $t$ is the Wannier downfolding nearest-neighbor hopping for the Ni 3$d_{x^2-y^2}$ band. (orange) Exchange interaction $J$ calculated by $4t^2/U$, where $t$ is estimated by $\frac{1}{8}\times$bandwidth of Ni 3$d_{x^2-y^2}$ band and $U$ is calculated by linear-response method \cite{Cococcioni2018_PRB}. (magenta) Cuprates exchange interactions are taken from Ref. \onlinecite{NdBCO, LCO, HgBCO, CCO}.}
\end{figure}
 
Let us quantify these superexchange  couplings. Considering either the effective bandwidth (Fig.~\ref{fig:figure3}a) or  the Wannier-downfolding (Table~\ref{tab:table1}) the kinetic scale $t_{\textrm{eff}}$ in the {\R}NiO$_2$ series is very similar to that in the cuprates.  The nature of the effective interaction provides a distinction.  The cuprate parent compounds are charge-transfer insulators with $U_{\textrm{eff}}\sim\Delta\sim3$ eV.  The {\R}NiO$_2$ series features a charge-transfer $\Delta\sim U$ and a reduced O-content in the lower Hubbard band, placing these compounds closer to  the single band Mott insulators, at least within the NiO$_2$-layer, such that $U_{\textrm{eff}}\sim U\sim6$ eV. Taking the standard perturbative expression $J=4\,t_{\textrm{eff}}^2/U_{\textrm{eff}}$ for the superexchange, the value in the NiO$_2$-layer would be smaller than that in cuprates, but only by approximately a factor of two, or even less.  To further quantify this comparison, we show the perturbative value of the superexchange $J$ across the lanthanide series in Fig.~\ref{fig:figure5}.  We use both the Wannier-downfolded nearest-neighbor hopping ($t^{Ni}_{[1,0,0]}$ from Table~\ref{tab:table1}) and one-eighth of the in-plane bandwidth to estimate $t_{\textrm{eff}}$. The later slightly overestimates the effective hopping due to longer range terms. To verify that using a static value of $U$ was a reasonable approximation in the estimate of $J$ across the lanthanides, we used a DFT perturbation linear-response method to calculate $U_{lr}$ for each $R$NiO$_2$ \cite{Cococcioni2018_PRB}. This calculation gave a higher average value of $U_{lr}=9.36$ eV but with a very small standard deviation of only 0.15 eV. As one can see from Fig.~\ref{fig:figure5}, a clear trend emerges across the lanthanide series in each case, with $J$ getting progressively larger. For comparison we show experimental estimates of $J$ in the cuprates for CaCuO$_2$,\cite{CCO} La$_2$CuO$_4$,\cite{LCO} HgBCO \cite{HgBCO}, and NBCO \cite{NdBCO}. The conclusion is that 
 the magnitude and nature of the superexchange interactions in the nickelate are predicted to be similar to the spin-spin interactions in the cuprates.

\section{Outlook}
One of the dimensions that can be used in the laboratory to find out which factors do matter for electronic properties like the superconductivity is chemical composition. 
In the case of the present monovalent nickelates preparing samples takes much effort to the degree that at the time of writing high quality bulk samples are not yet available. 
Band structure methods have evolved to a state that these are trustworthy with regard to predicting the large numbers as of relevance to the microscopic nature of the electron 
systems.  In this spirit we explored the role of the rare earth ions in the monovalent nickelates: is it worthwhile to substitute various rare-earth ions?

Given their chemical similarity we anticipate that such substitutions are rather easy to accomplish. However, this comes with a price: given the chemical similarity of the lanthanides 
not much may change in the electronic structure. The case in point is in the early history of the cuprates where it was discovered that the rare earth substitution had  a disappointing 
effect on the superconducting $T_c$'s in the YBCO system \cite{PHHor}.  But the monovalent nickelates are predicted to be 
in this regard quite different from the cuprates. In the latter, the ionic spacer layers may be viewed as lanthanide oxides; the lanthanide ions reside in caches formed from oxygen 
and upon changing the ionic radii of the former the oxygen packing has the effect that barely anything happens with the lattice constants. The monovalent nickelates are special
in the regard that all these oxygens are removed and the Ni-O perovskite layers are kept apart by just the lanthanide ions. As we showed, the outcome is that 
especially the c-axis lattice constant is shrinking substantially going from left to right in the series, by a magnitude similar to that found in the elemental rare earth metals. 

According to our calculations, this is an interesting ``knob" to tune microscopic parameters governing the electron system. Given the knowledge accumulated in the cuprates there is reason to suspect that these may well be important for matters like the superconducting transition temperature. However, in the cuprates these cannot be as neatly controlled as in the nickelates, given that there is no analogy of the ``lanthanide volume control knob" that we identified in the nickelates.

We find that the gross picture for the electronic structure of the monovalent nickelates is closely related to the cuprates. It consists of nearly decoupled two dimensional 
Hubbard-like systems living on a square lattice. Without doping these would form a Mott insulating, unfrustrated, Heisenberg antiferromagnet, becoming a ``Mottness" metal upon doping 
were it not for the fact that these layers are immersed in a low density three dimensional Fermi-liquid formed from rare earth 5$d$-like conduction band states. The latter 
are characterized by small electron-like Fermi pockets rendering the system metallic even at half filling, while these metallic electrons are subjected to a subdominant Kondo-like interaction 
with the strongly correlated Ni-O $d_{x^2-y^2}$ electrons. 

How can the ``lanthanide knob" be exploited to find out how this electron system works? A first expectation is that the 3D ``pockets" may be altered by the shrinking 
c-axis. However, we find that the influence of the c-axis on the pockets is rather marginal: the LP actually shrinks while another small pocket may spring into existence
pending the (unknown) value of $U$. Quite generally, one may turn it around to claim that measurement of this pocket, {\it e.g.} by photoemission or even quantum oscillations,  may be
quite informative regarding the details of the microscopic electronic structure that cannot be determined accurately on theoretical grounds (like the $U$). 

Since the very beginning \cite{Emory}, a debate has been raging in the high $T_c$ community that continuous until the present \cite{Adolphs} 
regarding whether one can get away with a single band Hubbard model or whether it is crucial to take into account the multi-band nature of the electronic structure, with prime 
suspects being the O 2$p$ states.  A complication is that the way in which the hole states are distributed over oxygen and copper in the various cuprate families is a rather complex 
affair, resting on chemistry that is not easy to control systematically. The gravity of this affair is highlighted in a recent study exploiting nuclear quadrupole resonance (NQR) to determine these charge distributions \cite{Rybicki, Jurkutat}, which shows a rather strong correlation between the magnitude of the $T_c$'s and the hole density on oxygen throughout the landscape of cuprate families.  In this regard, the monovalent nickelates may be a new theatre to study the issue. Is it as simple as $T_c$'s being lower because the oxygen band is at a higher charge transfer energy than in the cuprates, with the effect that these materials are closer to a ``single-band'' picture?  The main surprise of our study is the unanticipated strong dependence of the oxygen character of the holes as a function of the lanthanide, highlighted in Fig.~\ref{fig:figure3}a. As we already stressed, lanthanide substitution should be easy to accomplish
with the main effect, according to our computations, being to substantially change the oxygen character of the holes. This oxygen content  is in principle measurable (e.g., by NQR, core hole spectroscopies). Assuming that there is indeed the strong correlation between oxygen character and $T_c$ \cite{Rybicki}, the prediction follows that the optimal $T_c$  should decrease strongly upon substituting heavier lanthanides, given that according to our results the other microscopic electronic structure parameters should change only marginally.  Such a possibility awaits rigorous experimental trial.

\section{Methods}

Density functional theory GGA and GGA + $U$ calculations for {\R}NiO$_2$ were performed using the GGA method and the simplified version by Cococcioni and de Gironcoli \cite{DFTU}, as implemented in Quantum ESPRESSO (QE) \cite{QE}, employing PAW pseudopotentials with the $4f$ electrons placed in the core (except for La), explicitly to cope with the inability of current DFT functionals to properly treat the localized $4f$ electrons. Ytterbium was not included in this rare-earth series analysis because the available pseudopotentials force zero occupation of the 5$d$ orbitals, resulting in fewer total valence electrons treated explicitly in the DFT calculations. This difference led to drastic changes in the Yb bandstructure from the trends observed over the rest of the series. Cerium was not included in this rare-earth series analysis because of the possible mix-valency. The remaining rare-earth elements had identical pseudopotential valence configurations.

The self-consistant field (SCF) calculations used a k-point Monkhorst-Pack grid of $16\times 16\times 16$, the non-self consistent field (NSCF) calculation was run with 1000 evenly distributed k-points for PDOS. In the bandstructure calculations there were 20 k-points along all high-symmetry lines, except along $\Sigma-Y$ which was 10. For a better estimate of pocket size, in the PDOS calculations, a gaussian broadening of 0.340eV (0.025Ry) and an energy grid step of 0.0025eV were used.

In the two Ni BCT unit cell and corresponding BZ, we find an antiferromagnetic solution with wave vector $(\pi, \pi, \pi)$ leads to the lowest energy.
For each rare-earth element, the {\R}NiO$_2$ crystal structure was relaxed with an unrestricted variable cell relaxation as implemented in QE. These relaxations were done with the Hubbard $U$ turned off and no spin polarization. All bandstructures were calculated using each rare-earth element's relaxed crystal structure.

GGA + $U$ calculations for CaCuO$_2$ were performed using Vienna Ab initio Simulation Package (VASP) \cite{VASP}. Including a Hubbard $U$ leads to spin polarization with the lowest energy states supporting a $G$-type antiferromagnet [($\pi$, $\pi$, $\pi$) spin ordering], slightly more energetically favorable than other types of antiferromagnetism or ferromagnetic ordering.  Antiferromagnetism doubles the unit cell (as shown in the inset of Fig.~\ref{fig:figure1}), and folds the bands within a new body centered tetragonal (BCT) Brillouin zone (BZ). The DFT+ $U$ calculations as shown in this study were all performed with antiferromagnetic calculation with wave vector $(\pi, \pi, \pi)$ in two Cu BCT. 

As already announced, we placed the 4$f$ electrons in the core. Especially for DFT+ $U$ calculations, this saves quite some computational effort with the added benefit that the figures representing the bandstructure 
do not get obscured by the many 4$f$ bands which likely are irrelevant for our purposes. Such a practice is usually legitimate when dealing with rare earth ions in ionic salts. The controlling factor is whether   
the {\R} ions have a ``stable" valency. For instance, the trivalent Ce ion is characterized by a $4f^1$ occupancy of the f-shell. Given that the $f$-shell is ``hidden" in the core of the ion
it is characterized by a very large $U > 10$eV while the hybridization with the valence states is minute. The outcome is that this $f$-electron turns into a strongy localized 4$f$ spin that 
barely interacts with the other electrons. In most cuprates one finds that this rare-earth spin system does not interfere with superconductivity and other intertwined orders at typical transition temperatures, while it shows 
a transition to an ordered magnetic state at very low temperature driven by dipolar interactions. These 4$f$ states only become active dealing with a mixed valent situation. The 
tetravalent state may become energetically close to the trivalent state, implying for exampe in Ce that one encounters $4f^1 - 4f^0$ valence fluctuation. This requires 
fine tuning, but cannot be excluded beforehand. In the cuprate context a famous example is the  PrBa$_2$Cu$_3$O$_{6.5}$ system, turning into an antiferromagnetic
insulator due to the mixed valence of the Pr ion \cite{Soderholm}.

The oxygen ratio in total DOS was determined by integrating the projected density of states (PDOS) corresponding to the lower Hubbard band for $U$ = 6 eV. The PDOS used for this integration had a gaussian broadenning of 0.054eV (0.004Ry), much lower than those shown in either Fig.~\ref{fig:figure3} or the Appendix, Fig.~\ref{fig:app_figure1}. The integration used the trapezoidal method as implemented in Python 3's \verb|numpy.trapz| \cite{numpy1, numpy2}. The range of integration was chosen by taking zeros of the first derivative of the PDOS near the Fermi energy and near the first local minimum $(\sim -0.5eV)$, which effectively straddles the lower Hubbard band. 

The size of the rare-earth electron pocket, or the Fermi surface volume as shown in Fig.~\ref{fig:figure3}b, was determined by approximating the surface as an ellipsoid:  the first two principle semi-axes approximate a circle in-plane, whose lengths are given by the average distance along the $\Gamma$-$\Sigma$, $\Gamma$-Y, and $\Gamma$-X directions; the length of the third principal semi-axis comes from the distance along $\Gamma$-Z. We report the volume as a percentage of the total volume of the first BZ. A higher resolution k-point path near $\Gamma$ was used in determining the Fermi energy-crossings $k_{\textrm{F}}$ for the rare-earth bands along these high-symmetry cuts, with 200 k-points in each direction.

Using the same higher resolution data near $\Gamma$, the effective mass was calculated from the second derivative of a parabolic fit to the bandstructure at the Fermi energy $k_{\textrm{F}}$. For parabolic fitting, ten points along the band both above and below $k_{\textrm{F}}$ were chosen and then mirrored symmetrically about $\Gamma$ (40 points total per fit). These points were fit with a second order polynomial $ax^2+bx+c$ \cite{scipy}, such that $d^2E/dk^2=a \pm \delta a$, where $\delta a$ is the standard deviation of the fit parameter.


\acknowledgements

We thank Danfeng Li, Hye-ok Yoon, and Das Pemmaraju for insightful discussions. The work at Stanford University and SLAC National Accelerator Laboratory was supported by the US Department of Energy, Office of Basic Energy Sciences, Division of Materials Sciences and Engineering, under Contract No. DE-AC02-76SF00515. This research used resources of the National Energy Research Scientific Computing Center (NERSC), a US Department of Energy Office of Science User Facility operated under Contract No. DE-AC02-05CH11231.

\bibliography{biblio}

\clearpage

\appendix

\begin{figure*}[h!]
\section{Lanthanide Series}
\includegraphics[width=2\columnwidth]{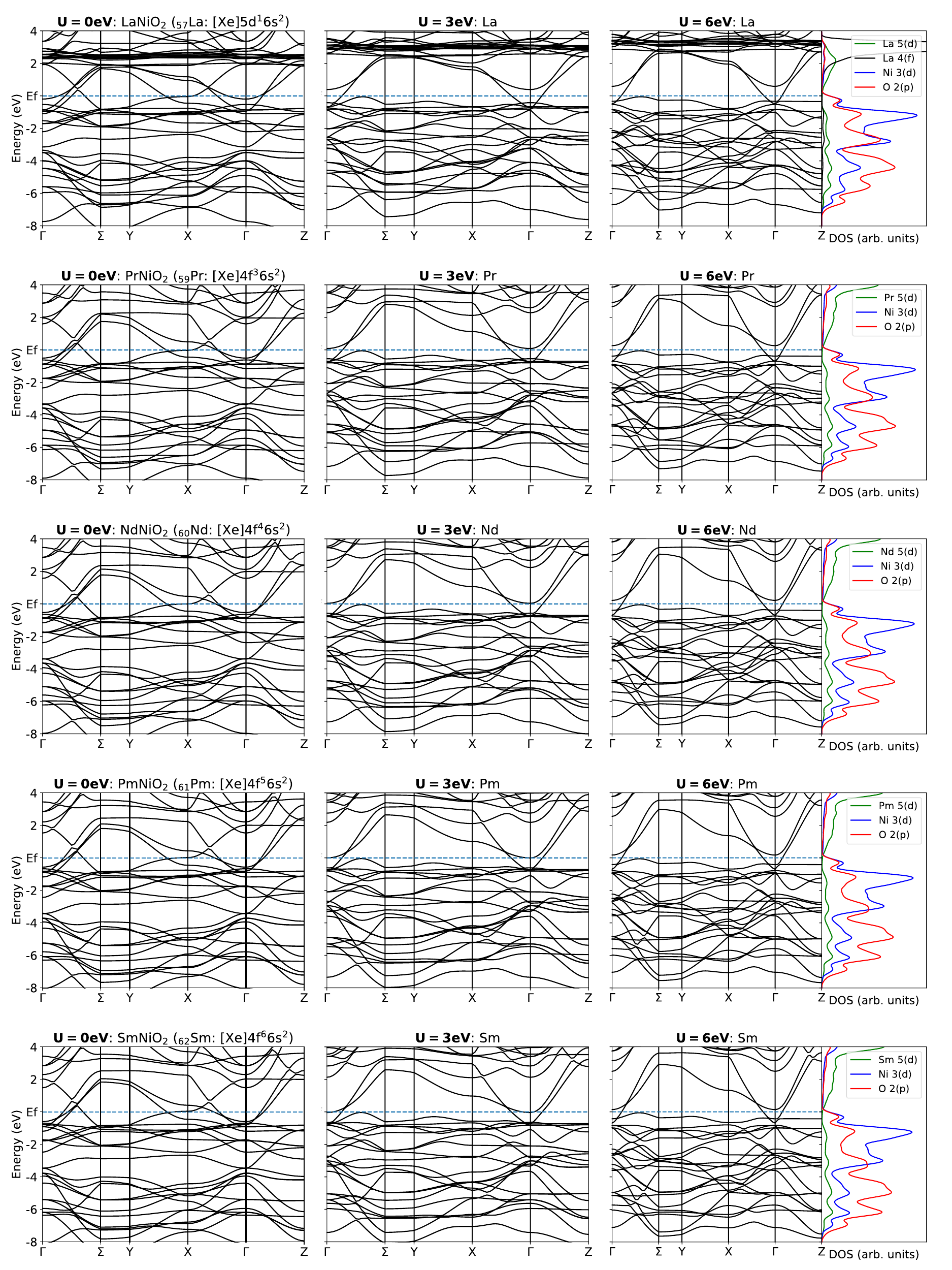}
\caption{\label{fig:app_figure1}  
{\textbf{Bandstructures of $R$NiO$_2$ calculated by GGA, GGA+$U$($U$=3 eV),GGA+$U$($U$=6 eV) and partial density of states of $R$NiO$_2$ calculated by GGA+$U$ (6 eV).}}}
\end{figure*}

\renewcommand{\thefigure}{\arabic{figure} (Cont.)}
\addtocounter{figure}{-1}

\renewcommand{\thefigure}{\arabic{figure}}

\begin{figure*}[h!]
\includegraphics[width=2\columnwidth]{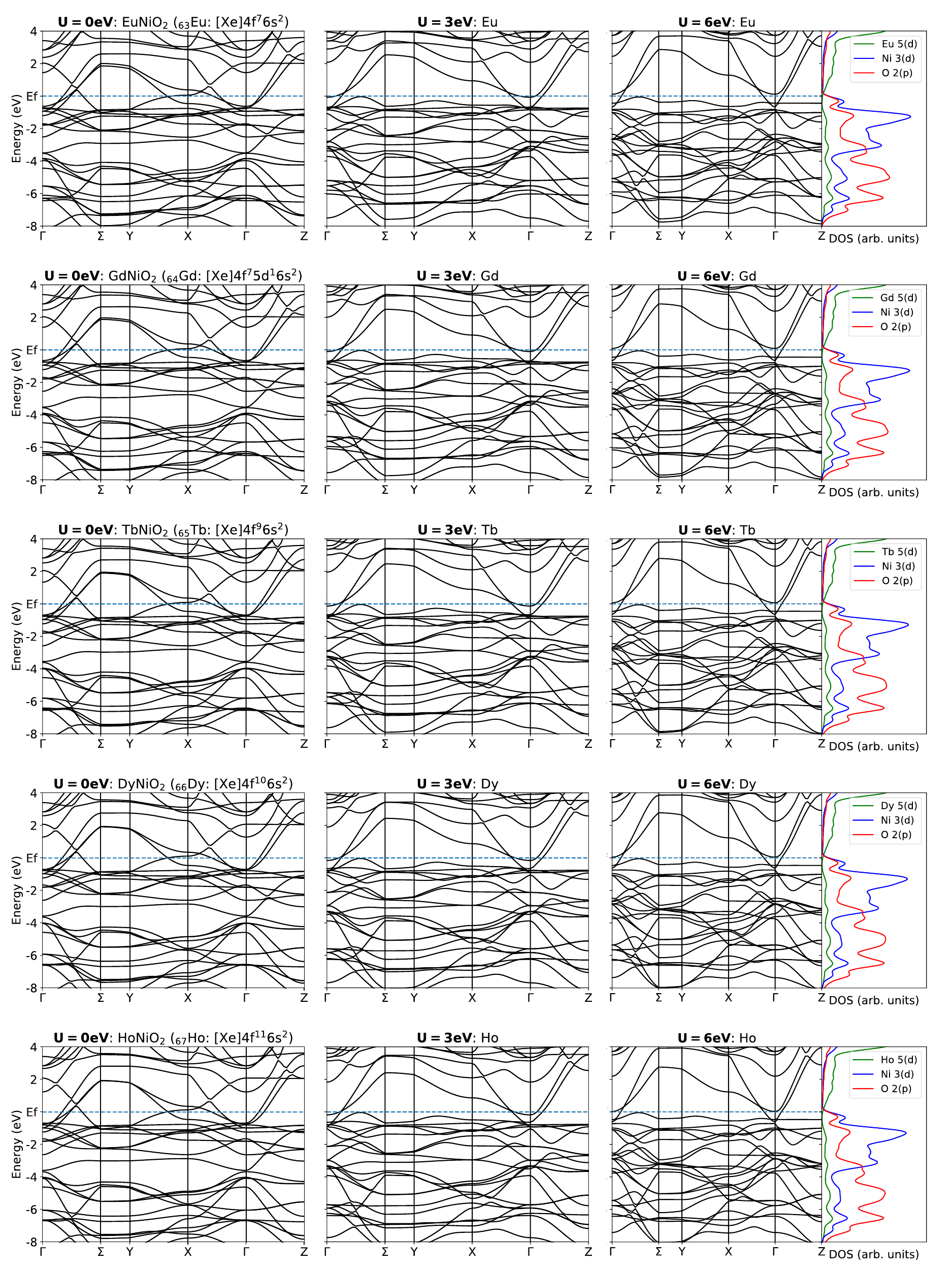}
\caption{\label{fig:app_figure1c}  
{\textbf{(continued) Bandstructures of $R$NiO$_2$ calculated by GGA, GGA+$U$($U$=3 eV),GGA+$U$($U$=6 eV) and partial density of states of $R$NiO$_2$ calculated by GGA+$U$ (6 eV).}}}
\end{figure*}

\addtocounter{figure}{-1}
\renewcommand{\thefigure}{\arabic{figure}}

\begin{figure*}[h!]
\includegraphics[width=2\columnwidth, trim={0 10cm 0 0}, clip]{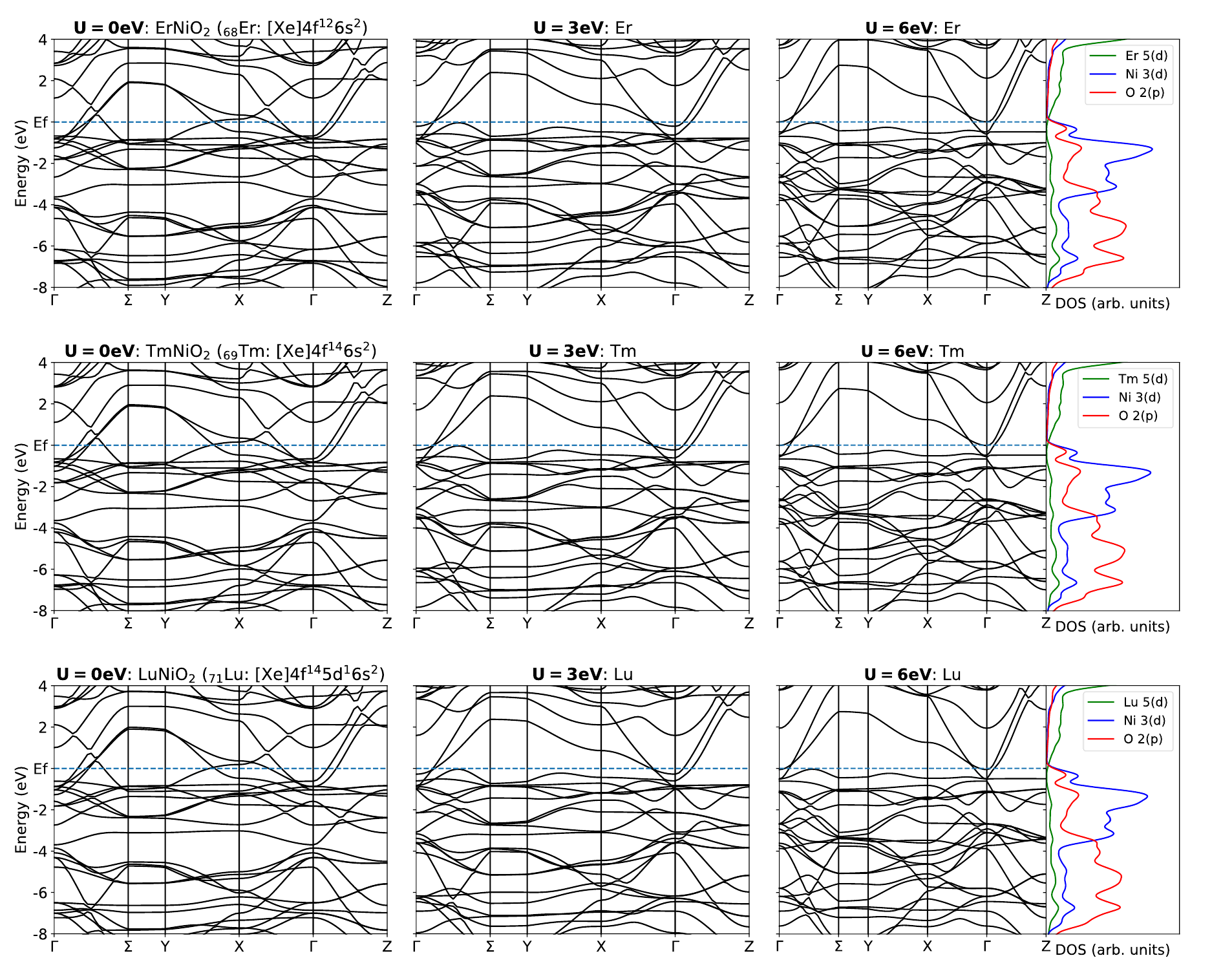}
\caption{\label{fig:app_figure1c}  
{\textbf{(continued) Bandstructures of $R$NiO$_2$ calculated by GGA, GGA+$U$($U$=3 eV),GGA+$U$($U$=6 eV) and partial density of states of $R$NiO$_2$ calculated by GGA+$U$ (6 eV).}}}
\end{figure*}

\begin{figure*}[h!]
\vspace{1mm}
\section{CaCuO$_2$ comparison bandstructure and PDOS}
\vspace{-2mm}
\includegraphics[width=1.5\columnwidth, trim={0 3cm 0 0}, clip]{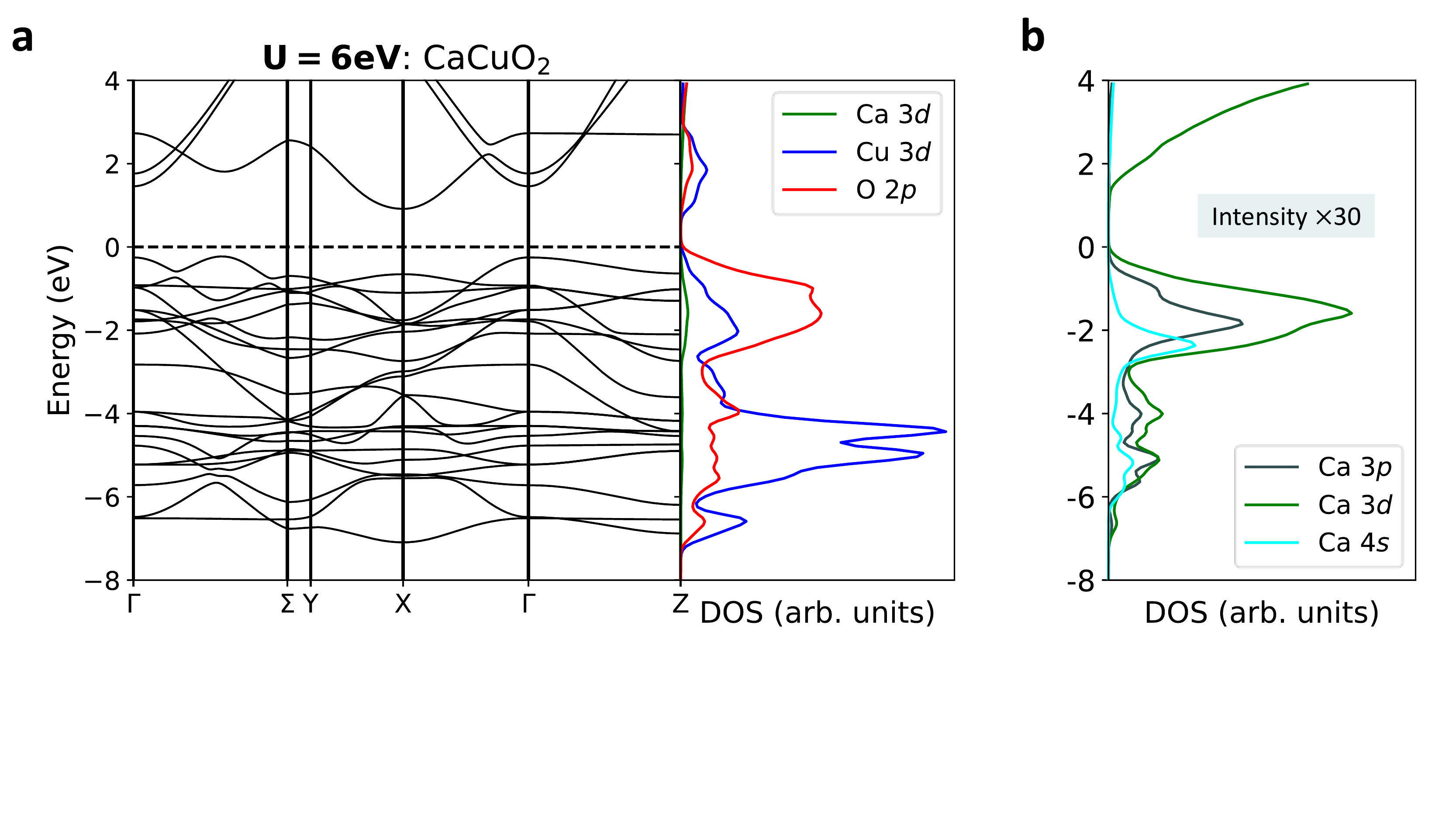}
\caption{\label{fig:app_figure2}  
{\textbf{Electronic structure calculation of CaCuO$_2$.}a, Bandstructures and partial density of states of CaCuO$_2$ calculated by GGA+$U$($U$=6 eV). b, Zoom-in intensity of Ca partial density of states of CaCuO$_2$ calculated by GGA+$U$ (6 eV).}}
\end{figure*}

\begin{figure*}[!h]
\section{Simple volume change with static rare-earth element $R=\text{Nd}$}
    \centering
\includegraphics[width=0.9\textwidth, trim={0 0 15cm 0},clip]{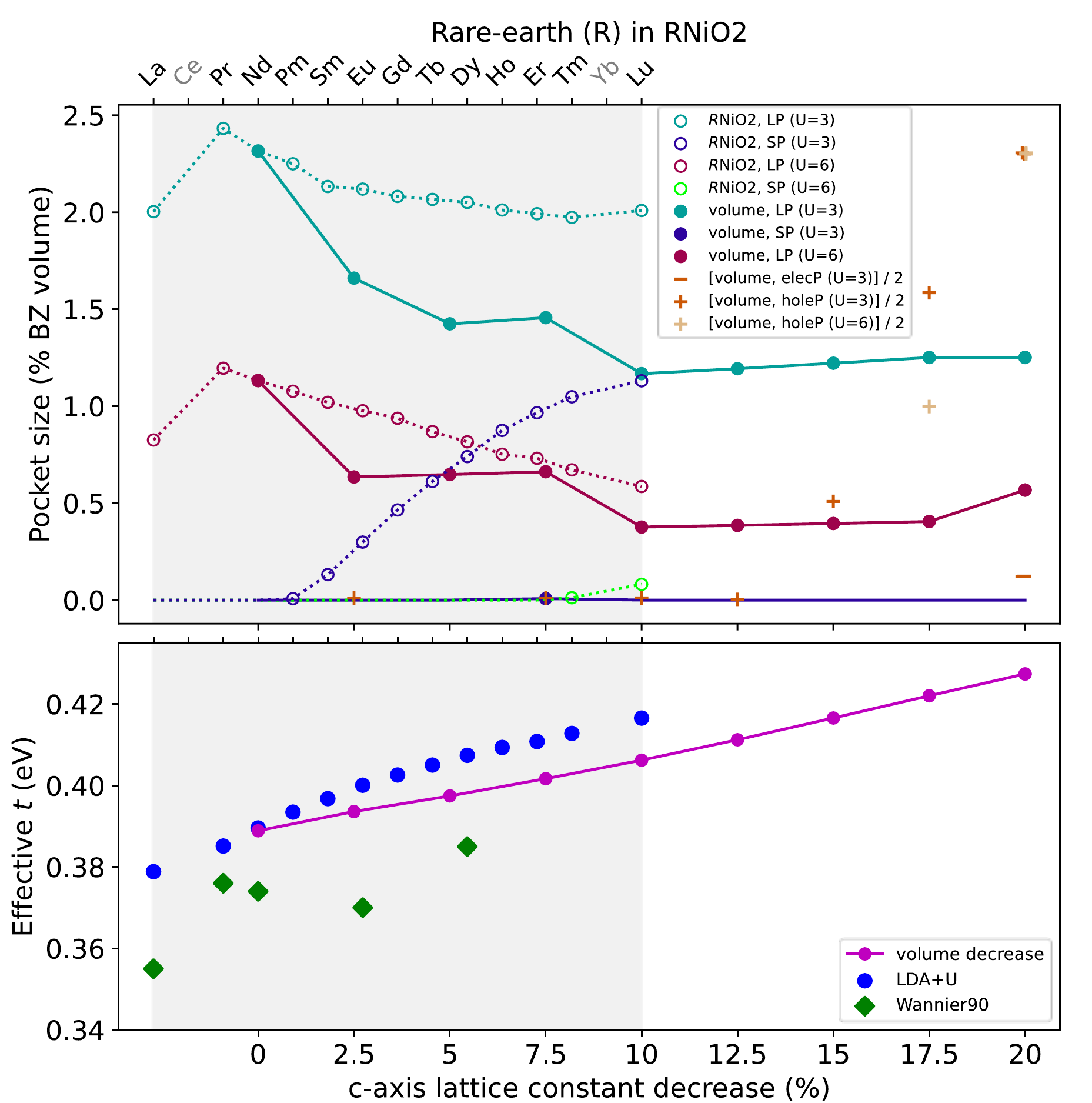}
        \caption{
        \textbf{Comparison between volume change with static $R=\text{Nd}$ and when rare-earth $R$ changes in $R$NiO$_2$}: (top) Effects on electron pockets for the two scenarios. LP and SP represents large pocket and small pocket for $R$ 5d orbital, labeled with circles. When only the volume changes and rare-earth stays constant, the second electron pocket (small pocket labeled SP) never substantially appears, and hole pockets start appearing in various parts of the BZ (namely along $Y-X$ and some along $\Gamma-\Sigma$). These anomalous hole pockets (and one electron pocket at 20\%) that appear along the BZ edge are shown at $0.5\times$ their actual values.
        (bottom) Hopping $t$ estimated by  $\frac{1}{8}\times($Ni $3d_{x^2-y^2}$ bandwidth) from the non-interacting DFT calculations. Changes in $t$ from exchanging rare-earch $R$ are significantly greater than those from a simple volume change.
        \\
        \\ \\
        When the volume change calculations were conducted, NdNiO$_2$ relaxed structure was used as equilibrium structure (labeled as 0\% decrease),  c-axis lattice constant decreased in 2.5\% increments ranging from 0\%-20\%, a- and b-axis lattice constants correspondingly decreased by 10\% of the c-axis lattice constant's decrease (e.g. when c decreased by 10\%, both a and b decreased by 1\% of their equilibrium values). These decreases are comparable to those seen in Figure \ref{fig:figure1} where there is a 10\% decrease in the c-axis lattice constant and 1\% decrease of the a- and b-axis lattice constants going from La to Lu. Although the 10\% decrease in c-axis lattice constant would correspond to the change across the lanthanides, we decided to extend this analysis to 20\% to see the effects of increased ``pressure" on the bandstructures.\\}
       
           \label{fig:simple_volume_change}
\end{figure*}
    
\begin{figure*}[h!]
    \section{Rationale to leave Ni $3d_{z^2}$ out of low-energy Hamiltonian}
    \centering
    \includegraphics[width=0.9\textwidth, trim={0 5cm 0 0}, clip]{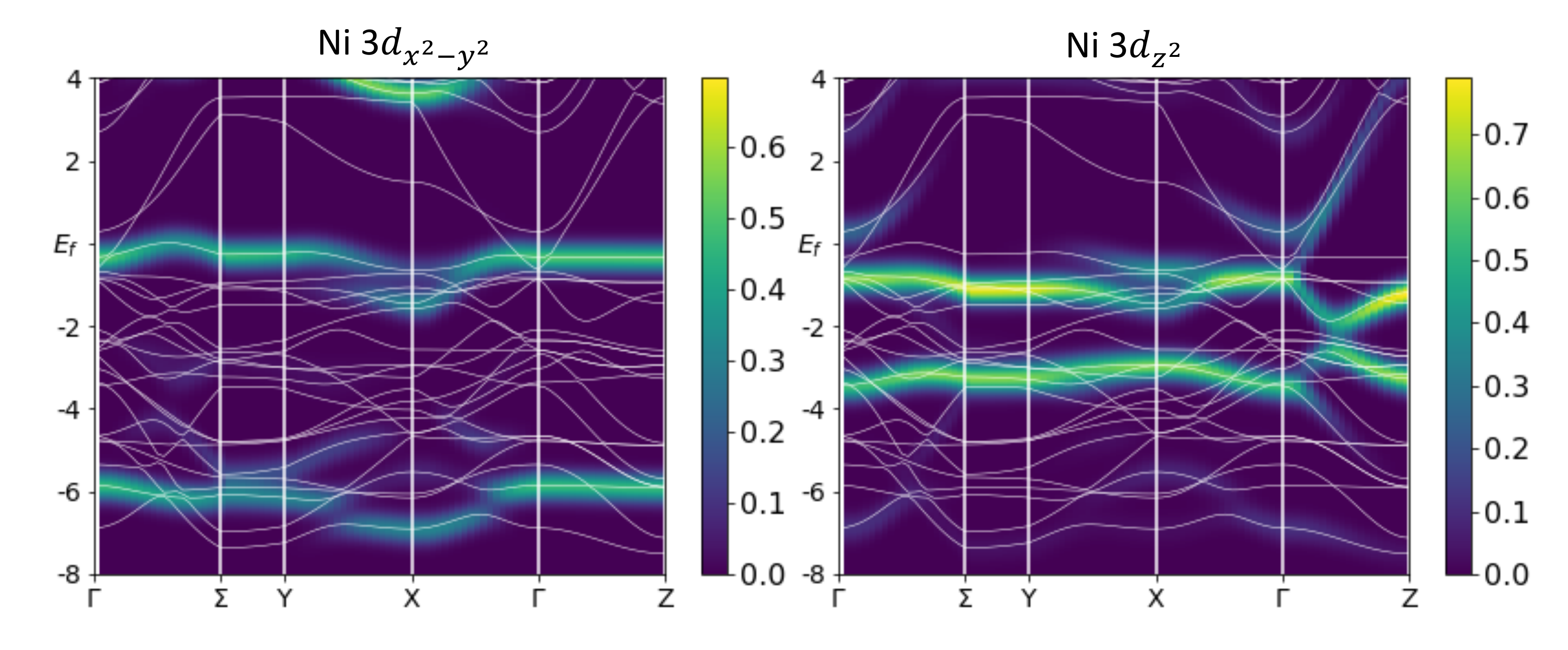}
    \caption{\textbf{Highlighted Ni $3d_{x^2-y^2}$ and $3d_{z^2}$ orbital character for the NdNiO$_2$ DFT+U ($U=6$ eV) calculation.}
    }
    \label{fig:color_boi}
\end{figure*}

\begin{figure*}[h!]
    \centering
    \includegraphics[width=0.5\textwidth, trim={0 5cm 15cm 0}, clip]{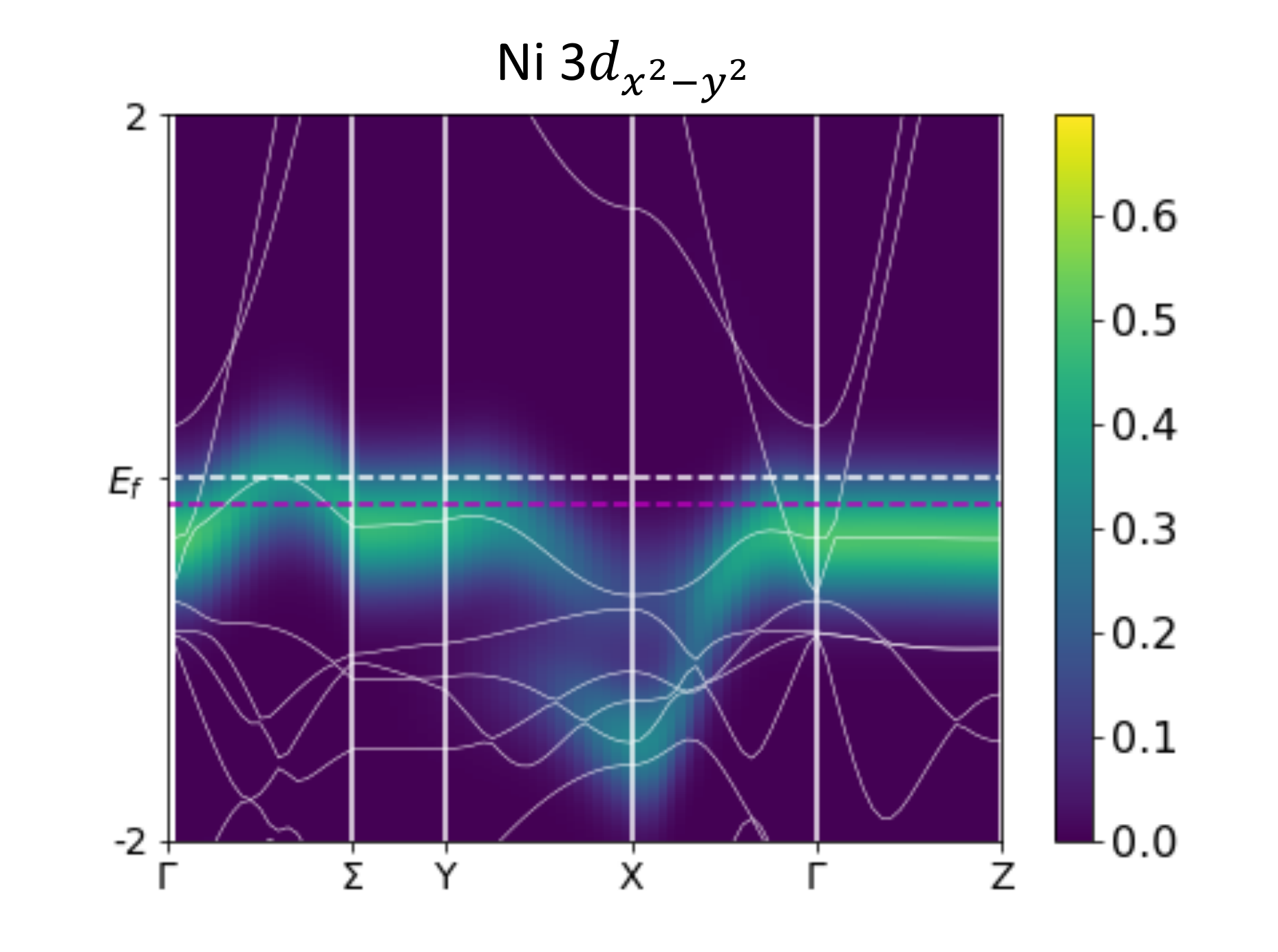}
    \caption{\textbf{Estimated Fermi energy shift for Nd$_{0.8}$Sr$_{0.2}$NiO$_2$}: $E_f$ for parent compound NdNiO$_2$ is marked in white dashed line; while estimated $E_f$ for  Nd$_{0.8}$Sr$_{0.2}$NiO$_2$ is marked in red dashed line.
    \\ \\ \\
    Consider the simplistic hypothesis that hole-doping applies a rigid shift to the Fermi level and leaves the rest of the bandstructure unchanged (although we know the Ni-derived bands will behave differently due to correlation effects, we use this simplistic ``doping" to get an estimate of which bands could potentially be affected). To approximate how far the Fermi energy will shift, we compare the valence electrons of the parent compound NdNiO$_2$ to the experimentally superconducting compound Nd$_{0.8}$Sr$_{0.2}$NiO$_2$ \cite{Danfeng2019}. In the NdNiO$_2$ compound, the pseudopotentials included 66 valence electons in the two-Ni unit cell. However, the total number of valence electrons in the pseudopotentials for Nd$_{0.8}$Sr$_{0.2}$NiO$_2$ is 65.6 electrons. Integrating the density of states for NdNiO$_2$ to this lower filling level gives us a Fermi Energy 0.14 eV lower than its current level. In the figure, this simplified estimate of the Fermi energy for Nd$_{0.8}$Sr$_{0.2}$NiO$_2$ is shown as the magenta dashed line, and the white dashed line denotes the Fermi energy for NdNiO$_2$ for Ni $3d_{z^2}$ in the NdNiO$_2$ DFT+U ($U=6$ eV) calculation. It is evident that even with the simplistic estimate of a rigid shift Fermi level relative to the bandstructure, all of the relevant information is still contained within the Ni $3d_{x^2-y^2}$ band since the $3d_{z^2}$ band is far below the ``doped" Fermi level.}
    \label{fig:color_boi_zoom}
\end{figure*}







\end{document}